\documentclass{jfm}

\graphicspath{{figures/}}

\usepackage{hyperref}
\usepackage{graphicx}
\usepackage{epstopdf, epsfig}
\usepackage{float}
\usepackage{url}
\usepackage{natbib}
\usepackage{amsmath,amssymb,amstext,mathtools}
\usepackage{scalerel}
\usepackage{stackengine,wasysym}
\usepackage{subcaption}
\usepackage{xcolor}

\def\k{\boldsymbol{k}}
\def\u{\boldsymbol{u}}
\def\v{\boldsymbol{v}}

\def\fU{\bar{\boldsymbol{u}}}
\def\fph{\bar{\phi}}
\def\fPh{\bar{\Phi}}

\def\D{\mathcal{D}}
\def\P{\mathcal{P}}
\def\x{\boldsymbol{x}}
\def\r{\boldsymbol{r}}

\usepackage{scalerel,stackengine}

\newcommand\reallywidehat[1]{\ensurestackMath{%
\savestack{\tmpbox}{\stretchto{%
\scaleto{%
\scalerel*[\widthof{\ensuremath{#1}}]{\kern-.6pt\bigwedge\kern-.6pt}%
{\rule[-\textheight/2]{1ex}{\textheight}}%WIDTH-LIMITED BIG WEDGE
}{\textheight}%
}{0.5ex}}%
\stackon[1pt]{#1}{\tmpbox}%
}}
\shorttitle{Dynamic Nonlocal LES Modeling for Scalar Turbulence}
\shortauthor{S. Hadi Seyedi, A. Akhavan-Safaei and M. Zayernouri}

\title{Dynamic Nonlocal Passive Scalar Subgrid-Scale Turbulence Modeling}

\author{S. Hadi Seyedi\aff{1},
\author,  Ali Akhavan-Safaei\aff{1,2}
 \and Mohsen Zayernouri\aff{1,3}\corresp{\email{zayern@msu.edu}}}

\affiliation{\aff{1}Department of Mechanical Engineering, Michigan State University, East Lansing, MI 48824, USA
\aff{2}Department of Computational Mathematics, Science and Engineering, Michigan State University, East Lansing, MI 48824, USA
\aff{3}Department of Statistics and Probability, Michigan State University, East Lansing, MI 48824, USA}

\begin{document}

\maketitle

\begin{abstract}
Extensive experimental evidence highlight that scalar turbulence exhibits anomalous diffusion and stronger intermittency levels at small scales compared to that in fluid turbulence. This renders the corresponding subgrid-scale dynamics modeling for scalar turbulence a greater challenge to date. We develop a new large eddy simulation (LES) paradigm for efficiently and dynamically nonlocal LES modeling of the scalar turbulence. To this end, we formulate the underlying nonlocal model starting from the filtered Boltzmann kinetic transport equation, where the divergence of subgrid-scale scalar fluxes emerges as a fractional-order Laplacian term in the filtered advection-diffusion model, coding the corresponding supper-diffusive nature of scalar turbulence. Subsequently, we develop a robust data-driven algorithm for estimation of the fractional (non-integer) Laplacian exponent, where we on-the-fly calculate the corresponding model coefficient employing a new dynamic procedure. Our \textit{a priori} tests show that our new dynamically nonlocal LES paradigm provides better agreements with the ground-truth filtered DNS data in comparison to the conventional static and dynamic Prandtl-Smagorisnky models. Moreover, in order to analyze the numerical stability and assessing the model’s performance, we carry out a comprehensive \textit{a posteriori} tests. They unanimously illustrate that our new model considerably outperforms other existing functional models, correctly predicting the backscattering phenomena at the same time and providing higher correlations at small-to-large filter sizes. We conclude that our proposed nonlocal subgrid-scale model for scalar turbulence is amenable for coarse LES and VLES frameworks even with strong anisotropies, applicable to environmental applications.
\end{abstract}

%\begin{keyword}
%Dynamic Modeling \sep Scalar Turbulence \sep Data-driven Modeling \sep Fractional order Calculus
%\end{keyword}

% =============================================================================================
\section{Introduction} \label{sec: Intro}
% =============================================================================================
Turbulence remembers and is fundamentally nonlocal. Such a longing portrait of turbulence originates from the delineation of coherent structures/motions, being spatially spotty, giving rise to interestingly anomalous spatio-temporal fluctuating signals \cite{davidson2015turbulence}.The statistical anomalies in such stochastic fields emerge as: sharp peaks, heavy-skirts of power-law form, long-range correlations, and skewed distributions, which scientifically manifest the non-Markovian/non-Fickian nature of turbulence at small scales \cite{ batchelor1953theory}.Such physical-statistical evidence highlights that `nonlocal features’ and `global inertial interactions’ cannot be ruled out in turbulence physics. On a whole different (computational) level and in addition to the aforementioned picture, the very act of filtering the Navier-Stokes and the energy/scalar equations in the large eddy simulations (LES) would make the existing hidden nonlocality in the subgrid dynamics even more pronounced, to which it induces an immiscibly mixed physical-computational nonlocal character. This urges the development of new LES modeling paradigms in addition to novel statistical measures that can meticulously extract, pin-down, and highlight the nonlocal character of turbulence (even in the most canonical flows) and their absence in the common/classic turbulence modeling practice. 

One of the oldest and most conventional local SGS modeling is known as the Prandtl-Smagorinsky model model (PSM) and was initially conceptualized in  \cite{smagorinsky1963general}. Despite being a significant step forward in LES studies, this eddy viscosity-based model suffers from a low correlation ratio, lack of back-scatter prediction, and flow-dependent features. Backward transfer of kinetic energy (back-scattering) from small scales to large scales is an innate feature of the turbulent flows that does reflect in the DNS and experimental studies. Nevertheless, most of the proposed turbulence models only predict the cascade of the energy from large to small scales. Another import weakness for the static PSM model is that there is not a single universal constant for the representation of different turbulent fields such as shear flows, rotating flows, or wall-bounded flows. As a remedy for the last two drawbacks, Germano et al. \cite{germano1991dynamic} proposed a new model, which is called the \textit{dynamic Prandtl-Smagorinsky model} (DPSM). They designed a dynamic procedure for the computation of the model constants as the calculation progresses. This procedure is based on the local calculation of the eddy viscosity coefficient by sampling the smallest resolved scales and using the obtained information in modeling the subgrid scales. Afterward, different dynamic models were designed and proposed based on the same concept \cite{piomelli1995large, najjar1996study, ghosal1997numerical}. 

Simulations of turbulent flows using DNS and experimental studies demonstrate that turbulence is intrinsically nonlocal \cite{wilczek2017emergence, akhavan2020anomalous}. Nonlocality of turbulence emerges as the sharp peaks, heavy-skirts, and skewed probability density function (PDF) in statistics of the velocity/scalar increments. Nevertheless, most of the turbulence models are based on Boussinesq's turbulent viscosity concept, which assumes the turbulent stress tensor to be proportional to the local mean velocity gradient. Bradshaw \citep{bradshaw1973agardograph} discussed that this assumption is not necessarily true everywhere and it does fail in some scenarios like curved surfaces. However, local models dominantly were utilized due to their easier implementations and absence of handy and feasible nonlocal models.

Introducing the nonlocality concept to the mathematical models can be done in different ways for a variety of applications. The most applicable and convenient one is based on using generalized-order derivatives. In the recent years, there have been remarkable studies in utilizing generalized-order derivatives including anomalous rheology \citep{suzuki2021data}, damage modeling \citep{suzuki2021thermodynamically} and many more that can be found in \citep{naghibolhosseini2015estimation, suzuki2021fractional, suzuki2021anomalous, kharazmi2019fractional, zhou2020implicit}. In turbulence modeling, there is also an emerging interest in recent years in the developments of nonlocal models. Egolf and Hutter \citep{egolf2017fractional, egolf2020nonlinear} introduced nonlocal models based on the Reynolds-averaged Navier–Stokes (RANS) coarse-grained technique. Some additional works in this area can be found in \citep{hamba2005nonlocal, chen2006speculative, epps2018turbulence}. In the LES turbulence modeling, a fractional Laplacian-based model was developed for the homogeneous isotropic turbulent (HIT) flows in \citep{samiee2020fractional}. They utilized L\'evy stable distributions in kinetic level and finally derived a nonlocal model that addresses the non-Gaussian statistics of the turbulent flows. Later they extended their modeling approach through developing a tempered fractional model using a tempered L\'evy stable distribution, which resulted in a promising performance in \textit{a priori} and \textit{a posteriori} tests \citep{samiee2021tempered}. A fractional eddy-viscosity-based model proposed in \citep{di2021two} and \textit{a priori} tests were performed for the HIT and turbulent channel flow as canonical test cases. On the nonlocal turbulence modeling for the wall-bounded turbulent flows \citep{keith2021fractional} proposed a class of turbulence model based on the fractional partial differential equations with stochastic loads. Additionally, one can consult with some preliminary related studies in \citep{milovanov2014mixed, ali2014theory, gunzburger2018analysis}. 

Modeling of the residual flux for the LES of conserved passive scalars (such as temperature) transported in turbulent flow medium has also been an important direction in the computational turbulence research, natural, and engineering applications. Due to the advective coupling with the turbulent velocity field, the fluctuations in passive scalar concentration field are known to be more intermittent and non-Gaussian compared to the velocity field \citep{shraiman2000scalar, warhaft2000passive, sreenivasan2019turbulent}. This behavior results in considerably stronger deviations of passive scalar statistical temporal records (turbulent intensity and dissipation) from their mean values (in a stationary turbulent regime) when we compare them to their counterparts in the turbulent velocity fields \citep[see e.g.,][]{donzis2005scalar, donzis2010resolution, portwood2021interpreting, akhavan2021nonlocal}. As a result, the residual scalar flux emerging in the governing equation for the LES of a turbulent passive scalar is naturally carrying a complex dynamics. Using a well-resolved filtered DNS dataset for the residual scalar flux, it has been shown that the subgird-scale dynamics has a ``statistically nonlocal'' nature that cannot be ruled out through conventional functional means of modeling such as those relying on the local eddy-diffusivity assumption \citep[see][Sec. 3]{akhavan2021data}. Therefore, using a detailed mathematical derivation starting Kinetic theory, \citep{akhavan2021data} obtained a fractional-order SGS model for scalar flux that successfully reproduced the nonlocal and non-Gaussian behavior of the residual scalar flux. Similar to Smagorisky model for passive scalars, the nature of their model is static in terms of the model coefficient and for its identification they relied on an \textit{a priori} regression approach with respect to the FDNS data. In practice, this \textit{a priori} regression step maybe found cumbersome and a dynamic procedure seems to be a proper modification to improve the generality of this model.    

In this work, we develop a new dynamic nonlocal passive scalar (DNPS) subgrid scale closure model. Both \textit{a priori} and \textit{a posteriori} assessments were performed on the model in order to test its performance. The results were compared with the conventional static and dynamic Prandtl-Smagorinsky models. As well as improving the performance of the static nonlocal passive scalar model (NPS) \citep{akhavan2021data}, the new model incorporates back-scatter prediction and does not require prior knowledge for the determination of the model constant. 

%\textcolor{red}{(scalar turbulence and its importance, quick recap on the paper and upcoming parts)}
The remaining portions of this research are arranged as follows: in section \ref{sec: Gov-Eqns}, we provide the governing equations and derivation of the proposed model. In section \ref{sec: Calibration} we elaborate on the importance of the fractional order and its identification method. Section \ref{sec: A Priori} is dedicated to the \textit{a priori} assessments and comparing the performance of the new model with the conventional PSM-based ones. Continuing the model performance tests, in section \ref{sec: A Posteriori}, we analyze the model performance and its numerical stability inside a solver. Finally, in section,\ref{sec: Conclusion}, we summarize the findings by a conclusion.

% ==============================================================================
\section{Development of Dynamic Nonlocal Passive Scalar Model (DNPS)}\label{sec: Gov-Eqns}
% ==============================================================================
~

\subsection{Governing equations}\label{subsec: Gov-Eqns}

Focusing on the incompressible flow regime and the transport of a conserved passive scalar (such as temperature field) in that flow medium, Navier-Stokes (NS) and Advection-Diffusion (AD) equations are the set of governing equations that constitute the dynamics \citep{pope2001turbulent}. In the Large-Eddy Simulation (LES) of turbulent transport a generic spatial filtering operator, $\bar{\cdot}$, is applied to the NS and AD equations returning the LES governing equations \citep[see e.g.,][]{sagaut2006large} as
\begin{eqnarray}
    \frac{\partial \fU}{\partial t}+\fU \cdot \nabla \fU &=&-\frac{1}{\rho}\,\nabla\bar{p}+\nu \, \Delta\fU-\nabla \cdot \boldsymbol{\tau}^R; \quad \nabla \cdot \bar{\u}=0,\label{eqn: F-NS} \\
    \frac{\partial \fPh}{\partial t}+\fU \cdot \nabla \fPh &=&\D \, \Delta\fPh-\nabla \cdot \boldsymbol{q}^R.\label{eqn: F-AD}
\end{eqnarray}
In these equations, $\u=(u_1,u_2,u_3)$, $p$, and $\Phi$ are the velocity, pressure, and scalar concentration fields, respectively. In \eqref{eqn: F-NS}, $\rho$ denotes the fluid density, while $\nu$ represents the viscosity of fluid, and in \eqref{eqn: F-AD}, $\D$ is the diffusivity of the passive scalar field. Moreover, filtering yields sources of closure in the LES governing equations as the divergence of \textit{residual stress}, $\boldsymbol{\tau}^R = \overline{\u \u} - \fU \fU$, and \textit{residual flux}, $\boldsymbol{q}^R=\overline{\u \Phi} - \fU \fPh$. Modeling these residual or subgrid terms using the filtered or resolved flow fields is an essential gateway returning a closed set of equations that are suitable for a predictive and numerically stable LES \citep{meneveau2000scale, sagaut2006large}.

The Reynolds decomposition for a general field such as scalar concentration, $\Phi = \langle \Phi \rangle + \phi$, where $\langle \Phi \rangle$ is the ensemble-averaged part of $\Phi$, and $\phi$ denotes its fluctuating part \citep{pope2001turbulent}. In our problem setting, we consider a homogeneous isotropic medium for velocity field; therefore, $\langle \u \rangle =0$. For the passive scalar field, we assume the fluctuations are statistically homogeneous while we consider an imposed ensemble-averaged gradient as $\nabla\langle \Phi \rangle = (0,1,0)$ \citep{overholt1996direct, akhavan2020parallel}. As a result, the filtered AD equation \eqref{eqn: F-AD} is rewritten as:
\begin{align}\label{eqn: F-AD-fluc}
    \frac{\partial \fph}{\partial t}+\fU \cdot \nabla \fph =-\bar{u}_2 + \D \, \Delta\fph-\nabla \cdot \boldsymbol{q}^R.
\end{align}
In \eqref{eqn: F-AD-fluc}, the residual scalar flux may be restated as: $\boldsymbol{q}^R=\overline{\u \phi} - \fU \fph$. The goal of our study is to focus on developing ``predictive'' and ``automated'' approaches for modeling $\boldsymbol{q}^R$ in a dynamic setting.

\subsection{Nonlocal modeling for the residual scalar flux}\label{subsec: Nonlocal-modeling}
An important element of a predictive modeling is the capability of the model to reproduce the main characteristics of the quantity that aimed to be predicted. In the LES of turbulent transport, nonlocality of the SGS dynamics requires a careful attention so that the prediction of the important statistical quantities such as resolved scalar variance, $\langle \fph^2  \rangle$, would be realistic over the course of a long-term simulation. In a comprehensive study by \citep{akhavan2021data}, using a rich filtered DNS (FDNS) data set, it has been illustrated that $\Pi = -\langle \boldsymbol{q}^R \cdot \nabla \fph \rangle$, (which is the SGS contribution to the time evolution of resolved-scale scalar variance) has a strong nonlocal behavior towards the larger filter sizes in way that:
($i$) the normalized probability distribution function of $\Pi$ exhibits heavier tails, ($ii$) the normalized two-point correlation function, $-\langle \boldsymbol{q}^R(\x) \cdot \nabla \fph(\x+\r) \rangle/\Pi$, yields higher values at a fixed shift value, $\r$, especially within the inertial-convective subrange. Moreover, they showed that classical eddy-diffusivity modeling (EDM) for the SGS scalar flux fails to address this nonlocal behaviour regardless of the filter size.

As a result, \citep{akhavan2021data} developed a fractional order SGS model for the residual scalar flux that successfully reproduced the nonlocal behavior they observed in the filtered DNS data set. Their mathematical modeling was originated from investigating the source of LES closure at the kinetic level from the filtered Boltzmann transport equation (FBTE),  
\begin{equation}\label{eqn: FBTE}
\frac{\partial \bar{g}}{\partial t} + \boldsymbol{v}\cdot \nabla \, \bar{g} = -\frac{\bar{g}-\overline{g^\mathrm{eq}(\mathcal{L})}}{\tau_g}.
\end{equation}
FBTE \eqref{eqn: FBTE} governs the time-evolution of distribution function for a single passive scalar particle , $g=g(t,\x,\v)$ at time $t$ with particle's spatial location $\x$ and velocity $\v$. In \eqref{eqn: FBTE}, the well-know Bhatnagar–Gross–Krook (BGK) kinetic model for the collision of two particles is utilized and it is characterized with a single parameter called \textit{relaxation time} ($\tau_g$). Moreover the BGK model assumes a \textit{local} equilibrium distribution function $g^\mathrm{eq}(\mathcal{L})$ (known as the Maxwell distribution) for the two-particle collision, which is a normally distributed function of $\mathcal{L}$ that is parameterized by the locally conserved quantities. In the FBTE, the source of closure stems from the fact that the filtering operator does not commute with the collision operator, which results in $\overline{g^\mathrm{eq}(\mathcal{L})} \neq g^\mathrm{eq}\bar{(\mathcal{L})}$. Given the fact the $\overline{g^\mathrm{eq}(\mathcal{L})}$ is a filtered normal distribution and has a multi-exponential nature, \citep{akhavan2021data} modeled its behavior with an $\alpha$-stable L\'{e}vy distribution as a function of $\bar{\mathcal{L}}$, which closes the FBTE. Using the ensemble-averaging over the 3-D space, they derived the continuum-level filtered AD equation with the modeled residual scalar flux as:
\begin{equation}\label{eqn: SGS-flux-NPS}
    \boldsymbol{q}^{R} = -\mathcal{D}_\alpha \, \boldsymbol{\mathcal{R}} (-\Delta)^{\alpha-1/2} \, \bar{\phi}, \quad \alpha \in (0,1].
\end{equation}
In \eqref{eqn: SGS-flux-NPS}, $\boldsymbol{\mathcal{R}}(-\Delta)^{\alpha-1/2}(\cdot)$ represents the fractional order gradient operator through the Riesz transform (see the Appendix \ref{sec: Appendix1}), and $\mathcal{D}_\alpha$ is a positive real-valued model coefficient. For more details on the derivation, the interested readers are referred to \citep[][Sec. 4 and Appendix B]{akhavan2021data}.

In particular, obtaining $\D_\alpha$ requires \textit{a priori} model identifications such as regression utilizing the true values of $\boldsymbol{q}^R$ from filtered DNS data. This procedure, is inherently making the modeling procedure impractical and more complicated. In order to address this issue and elevate the modeling framework to an automated level, we develop a \textit{dynamic} procedure based on the fractional order SGS scalar flux given in \eqref{eqn: SGS-flux-NPS}.

\subsection{Dynamic nonlocal modeling for the residual scalar flux}\label{subsec: SGS-modeling}

Following the concept introduced by \cite{germano1991dynamic}, the fluxes at test-level filter (subtest-scale fluxes) can be written as
\begin{equation}\label{test-scale_flux}
    Q^{R}_{i} = -\mathcal{D}_\alpha \, \boldsymbol{\mathcal{R}}_i (-\Delta)^{\alpha-1/2} \, \widehat{\bar{\phi}}.
\end{equation}
where, $\widehat{(\cdot )}$ assigned for showing filtering at the test-level. The ratio between the test-level and grid-level filter sizes are usually chosen equals to two. Resolved scales and subgrid-scales are being related through the Germano identity.
\begin{equation}\label{Germano}
          G_{i} = Q^{R}_{i} - \widehat{q^{R}_{i}}, 
\end{equation}
replacing the previously achieved relation in \eqref{Germano} gives
\begin{equation}\label{Germano2}
          G_{i} = -\mathcal{D}_\alpha \, \boldsymbol{\mathcal{R}}_i (-\Delta)^{\alpha-1/2} \, \widehat{\bar{\phi}} + \reallywidehat{\mathcal{D}_\alpha \, \boldsymbol{\mathcal{R}}_i (-\Delta)^{\alpha-1/2} \, \bar{\phi}}.
\end{equation}
By assuming scale-invariance condition for the model constant and considering the known quantities for the definition of this identity, we can simplify the Germano identity as
\begin{equation}\label{Germano3}
          \reallywidehat{\bar{\phi} \, \bar{u}_i} - \widehat{\bar{\phi}} \, \widehat{\bar{u}}_i  = -\mathcal{D}_\alpha \, \bigg(  \boldsymbol{\mathcal{R}}_i (-\Delta)^{\alpha-1/2} \, \widehat{\bar{\phi}} - \reallywidehat{\, \boldsymbol{\mathcal{R}}_i (-\Delta)^{\alpha-1/2} \, \bar{\phi}}\bigg).
\end{equation}
It is a common assumption that the filtering procedure commutes with the integer-order derivatives and this assumption is basically one of the main steps of deriving the filtered equations in LES methods. However, we show that this is not a valid premise when one deals with the fractional operators. It can be seen in \eqref{Germano3} that the filtering process does not commute with the fractional operators. The difference between the fractional derivative of filtered scalar field and the filtered fractional derivative at the test-level filter is not a negligible amount. Actually, we are using this difference to estimate the behavior in the modeled section. Several methods have been suggested in this section for the calculation of the model constants. Since the scope is having a scalar model constant at the end, we need to first contract the tensorial quantities. One approach that was utilized originally by Germano \textit{et al.} \citep{germano1991dynamic} for the dynamic Smagorisnky method is based on contracting using the filtered strain rate tensor. The other usual method which is more robust and fruitful is suggested by \citep{lilly1992proposed}. The second approach is based on the least square method (LSM) and assumes a squared error as
\begin{equation}\label{Error}
     \bigg(  \reallywidehat{\bar{\phi} \, \bar{u}_i} - \widehat{\bar{\phi}} \, \widehat{\bar{u}}_i + \mathcal{D}_\alpha \, \mathcal{M}_i \bigg)^2 = e.
\end{equation}
in which $\mathcal{M}_i $ is defined as
\begin{equation}\label{M_i}
     \mathcal{M}_i = \boldsymbol{\mathcal{R}}_i (-\Delta)^{\alpha-1/2} \, \widehat{\bar{\phi}} - \reallywidehat{\, \boldsymbol{\mathcal{R}}_i (-\Delta)^{\alpha-1/2} \, \bar{\phi}} .
\end{equation}
Now, we can put the derivative of squared error equals zero to find the model constant. Also, we double-check for the correctness of  $\frac{\partial ^ 2 e}{\partial C^2} >0 $ in order to get the minimized values for the error. Therefore, the scalar model constant would be calculated as 
\begin{equation}\label{eq:C }
           \mathcal{D}_\alpha  = - \frac{\langle ( \reallywidehat{\bar{\phi} \, \bar{u}_i} - \widehat{\bar{\phi}} \, \widehat{\bar{u}}_i) \ \mathcal{M}_i  \rangle} {\langle   \mathcal{M}_i \ \mathcal{M}_i  \rangle}. 
\end{equation}
We also added averaging operators to avoid numerical instability due to the negative eddy diffusivity. This approach of averaging over the directions of statistical homogeneity is proposed by \citep{germano1991dynamic}. This paper focuses on the development of a dynamic, functional nonlocal model. The same concept can also be applied to build a hybrid model (functional + structural), but that is not the focus of this paper. In Appendix \ref{sec: Appendix1}, we have provided the necessary mathematical concepts and definitions that being used in the model development procedure. 

% =============================================================================
%                Numerical Results
% =============================================================================
\section{Data-driven Identification of Optimum Fractional Order, $\alpha^{opt}$}\label{sec: Calibration}

We introduce a data-driven approach to determine the optimum fractional order in the derived DNPS closure model. The new nonlocal proposed model, like every other nonlocal one, has a tuning knob which is called fractional order ($\alpha$). According to \eqref{eqn: SGS-flux-NPS}, this value can take values between zero and one, and setting different fractional orders would result in different fractional derivatives. Therefore, an optimization algorithm is needed to determine this value before deployment of the model in \textit{a priori} and the \textit{a posteriori} tests. Several approaches and criteria could be imagined for this optimization procedure such as (1) selection of $\alpha$ when gaining maximum correlations for the average of the scalar fluxes and ground-truth DNS results in three directions or (2)  when there is a maximum correlation for the divergence of fluxes from the model with the ground-truth results, and (3) when we have a minimum mean squared error comparing the divergence of the fluxes (closure terms). Preliminary analysis showed that the second and third approaches are providing very close results and also we have a better prediction of back-scatter phenomenon in case we use the second (or third) approach rather than the first one. Therefore, the second approach is selected and utilized in this research. 

We construct our database based on 10 sample snapshots of DNS simulations using the pseudo-spectral parallel code elaborated in \citep{akhavan2020parallel}. Using the mentioned framework we generate a stationary HIT flow with $520^3$ resolution. The computational domain is a cube as  $\boldsymbol{\Omega} = [0, 2 \pi]^3$, and the Taylor-scale Reynolds ($Re_\lambda$) and Schmidt ($Sc=\nu/\D$) numbers are $240$ and $1$, respectively. The 10 snapshots are sampled out from the DNS simulation after being sure of reaching statistically stationary condition over 10 large-eddy turnover times \citep{akhavan2020parallel, akhavan2021data}. Thereafter, the discussed optimization procedure is implemented on each snapshot and finally, the ensemble-averaged quantities are reported and utilized for the discussions and analysis. As it has been reported in \citep{seyedi2022data, samiee2021tempered,samiee2020fractional,akhavan2021data}, changing the LES filter size would change the optimum fractional order. Moreover, $\mathcal{L}_{\delta} =  \Delta^{*} \Delta_{DNS}$, where $\Delta^{*}$ is the ratio between the LES and DNS grids, and the DNS grid size is defined as $\Delta_{DNS} = \frac{2 \pi}{N}$ , $N=520$ is the DNS resolution.  In this study, we utilize three different filter sizes of ${\mathcal{L}_{\delta}} =4, 10 ,20$. The first filter size (${\mathcal{L}_{\delta}} = 4$) would represents a conventional LES filter size, the second filter size (${\mathcal{L}_{\delta}} = 10$) is called the coarse LES, and finally the third filter size (${\mathcal{L}_{\delta}} = 20$) is trying to test the model performance in the very large-eddy simulation (VLES) studies. Our criteria for this classification are based on the amount of filtered scalar variance that is being modeled in the turbulence modeling procedure. The amount of the modeled filtered scalar variance for these three filter sizes are $ 4\%, 9\%$ and $18\% $, respectively. 
\begin{figure}
    \begin{center}
        \begin{minipage}[b]{.49\linewidth}
            \centering
            \includegraphics[width=1\textwidth]{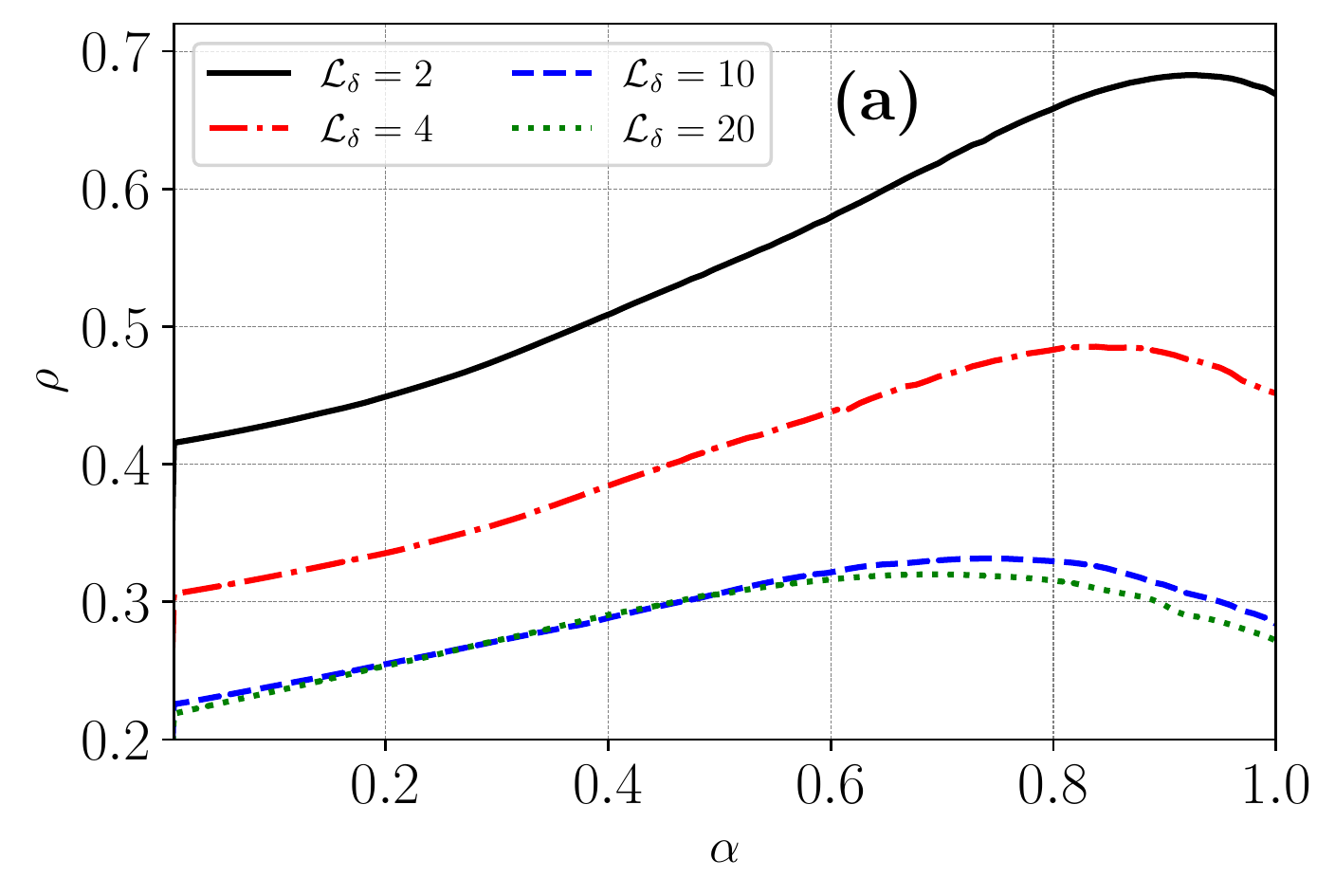}
        \end{minipage}
        \begin{minipage}[b]{.49\linewidth}
            \centering
            \includegraphics[width=1\textwidth]{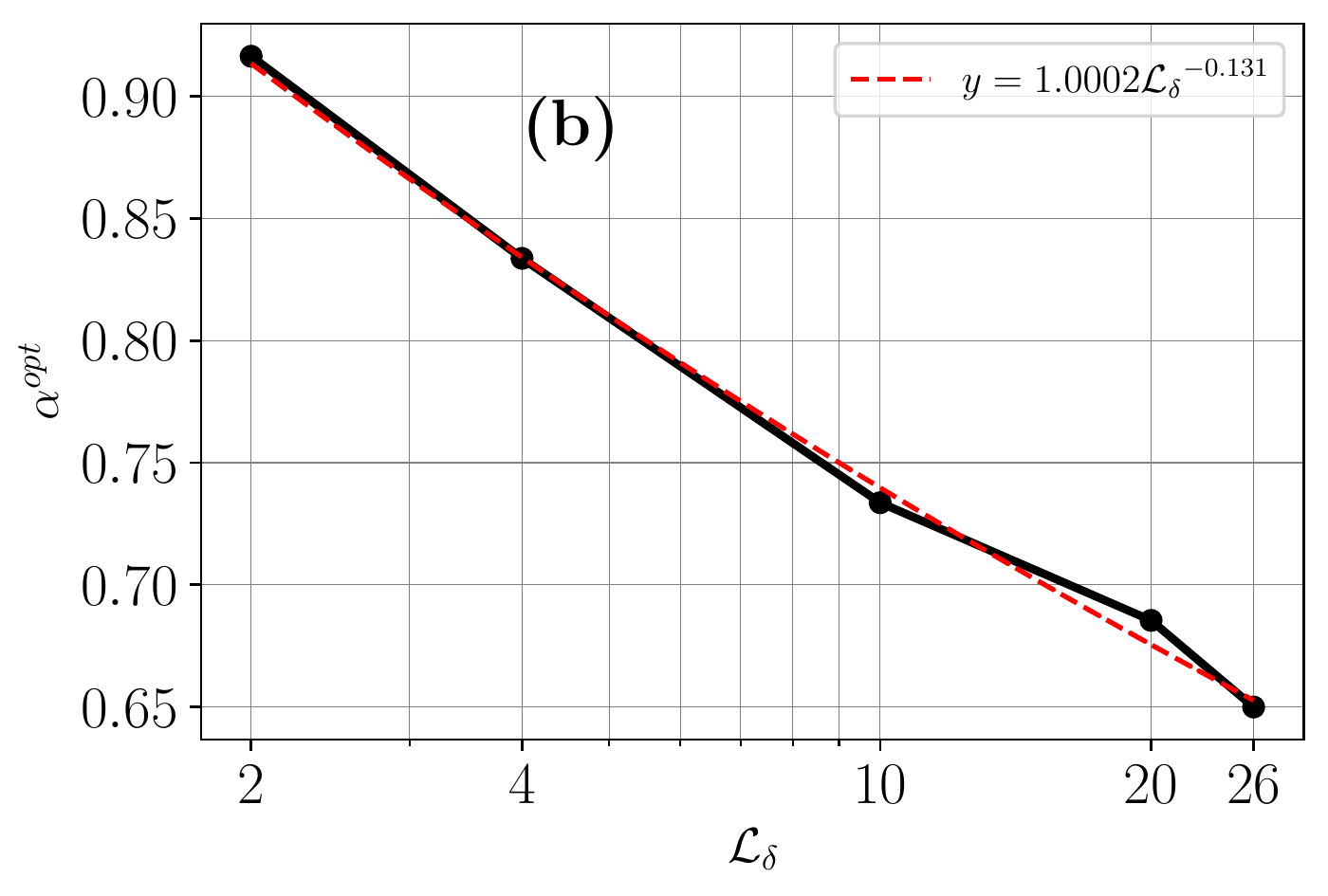}
        \end{minipage}
        \caption{\footnotesize  Obtaining $\alpha ^{opt}$ at different filter sizes in one of the snapshots of data and (b) Power-law relation between the $\alpha ^{opt}$ and ${\mathcal{L}_{\delta}}$ based on the ensemble-averaged results.}\label{fig: alpha_opt_all}
    \end{center}
\end{figure}
Figure \ref{fig: alpha_opt_all}(a) shows the process of finding optimum fractional order based on the highest correlation for the forces in different filter sizes in one of the sample snapshots of data. As it is clearly seen, the maximum values of the plots are moving toward zero by increasing the filter size. This reverse relation has also been reported in \citep{seyedi2022data, samiee2020fractional} and is due to the fact that by increasing the filter size, more nonlocality is incorporated (see section \ref{subsec: Nonlocal-modeling}). Increasing the filter size, usually results in lower correlation coefficients ($\rho$); however, one can see that by going from ${\mathcal{L}_{\delta} = 10}$ to ${\mathcal{L}_{\delta} = 20}$ there is not a significant reduction in the performance. This interesting feature will be elaborated more in the upcoming sections. Finding the exact values of the optimum fractional orders based on the ensemble averaging over the 10 snapshots of filtered DNS data, demonstrates that there is a power-law relation between the filter size ${\mathcal{L}_{\delta}}$ and $\alpha^{opt}$. As it has depicted in Figure \ref{fig: alpha_opt_all}(b), the trend-line is clearly showing a power-law behavior in which the $R^2$ values, the coefficient of determination, for these regression procedures are above $0.996$.

% =============================================================================
\section{\textit{A Priori} Tests }\label{sec: A Priori}
% =============================================================================
We perform a comprehensive \textit{a priori} test on the newly developed DNPS model and compare the results with the results of the conventional eddy-diffusivity-based model including static Prandtl-Smagorinsky (PSM) and dynamic Prandtl-Smagorinsky (DPSM) models. The ground-truth results are achieved using the filtered DNS data (FDNS).   

\begin{figure}
    \begin{center}
        \begin{minipage}[b]{.49\linewidth}
            \centering
            \includegraphics[width=1\textwidth]{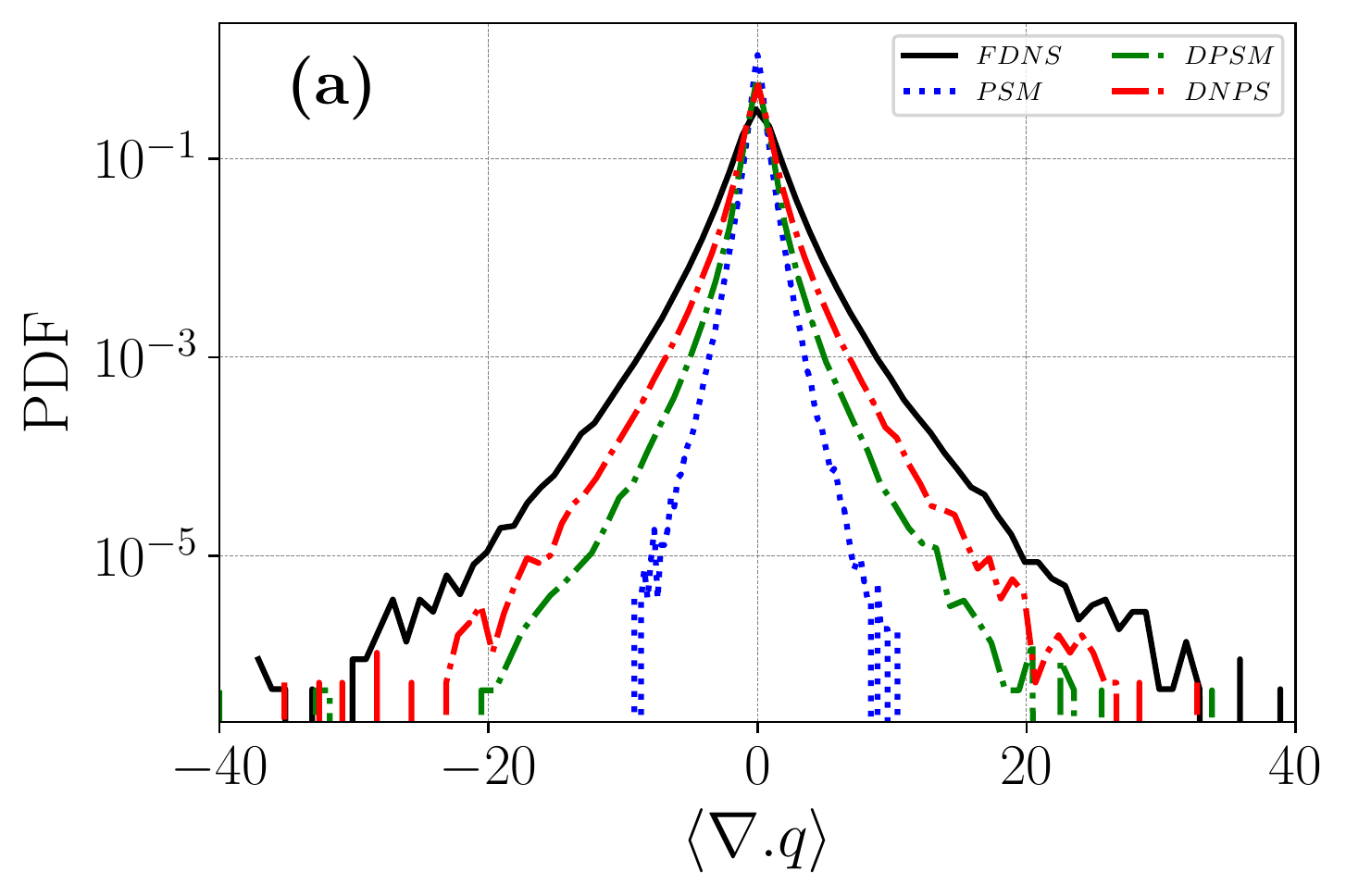}
        \end{minipage}
        \begin{minipage}[b]{.49\linewidth}
            \centering
            \includegraphics[width=1\textwidth]{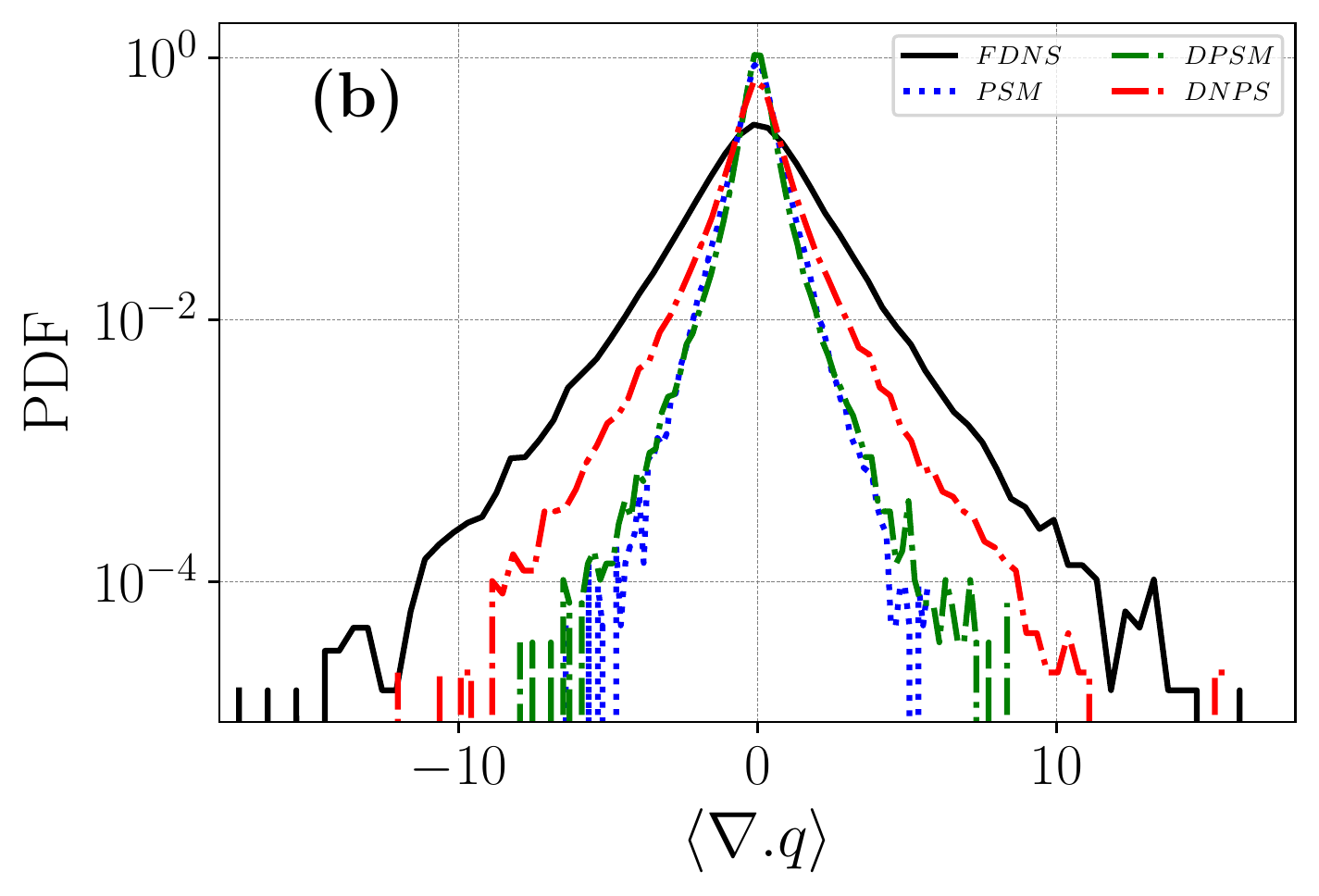}
        \end{minipage}
        \begin{minipage}[b]{.59\linewidth}
            \centering
            \includegraphics[width=1\textwidth]{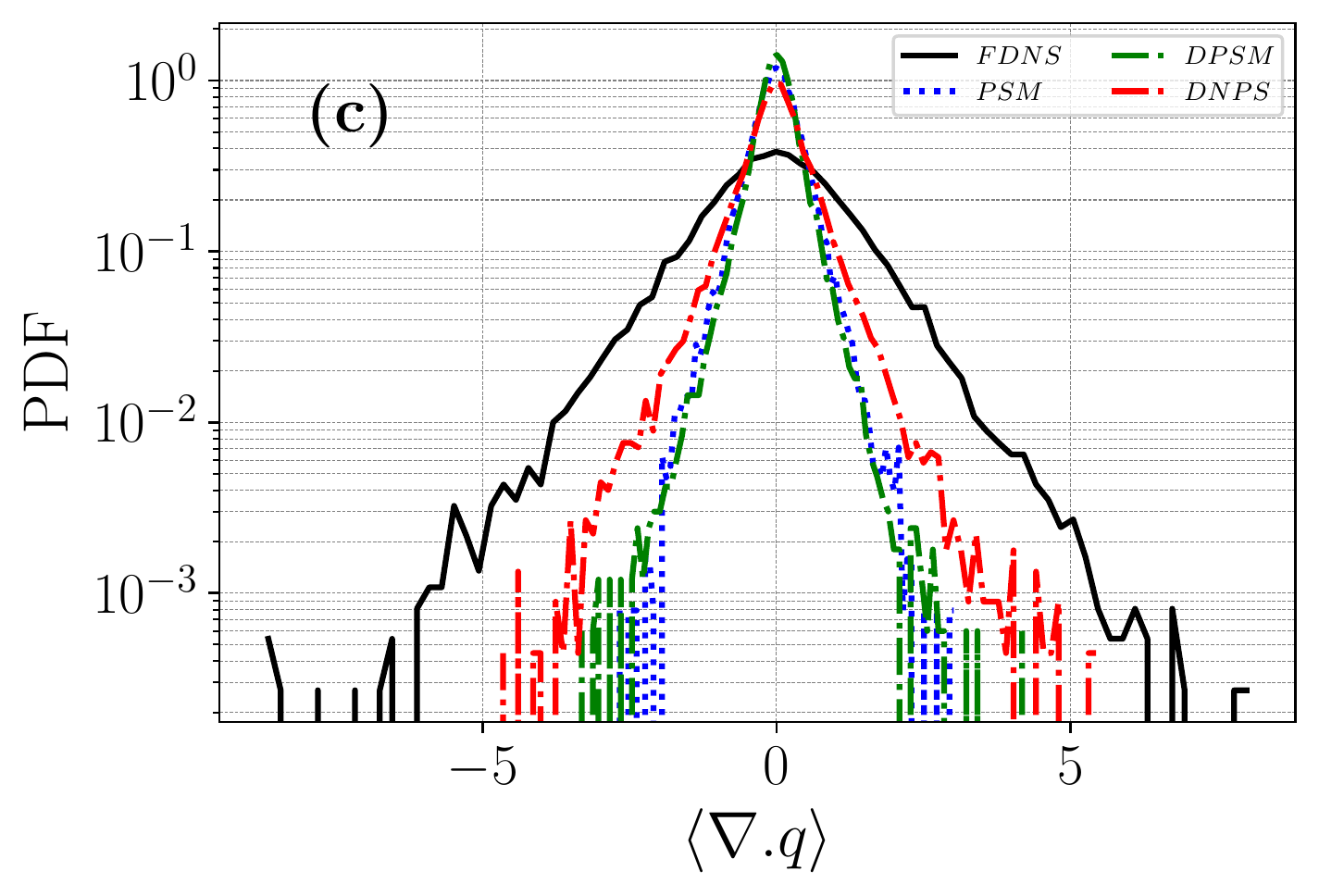}
        \end{minipage}
        \caption{\footnotesize  Normalized PDF of ensemble-averaged SGS forces using different models and different filter sizes, (a) ${\mathcal{L}_{\delta}} = 4$ for LES, (b) ${\mathcal{L}_{\delta}} = 10$ for coarse LES, and (c) ${\mathcal{L}_{\delta}} = 20$ for VLES.}\label{fig: Force}
    \end{center}
\end{figure}

We have depicted the ensemble-averaged forces related to the closure term in Figure \ref{fig: Force} using previously mentioned filter sizes to cover all scenarios regarding the characteristic LES filter size. The first graph (a) which belongs to the conventional LES scenario, ${\mathcal{L}_{\delta}} = 4$, shows that the new DNPS model provides better prediction in capturing the peak and both right and left wings of the PDF. Also, the one-point correlation coefficient between the DNPS and FDNS results is 0.48 which is remarkably higher than PSM and DPSM models with 0.19, 0.40 correlation coefficient, respectively.
Increasing the filter size to ${\mathcal{L}_{\delta}} = 10$ in second plot (b) clearly shows that conventional PSM-based models are deviating from the ground-truth values. In the third plot (c), we have the results for the ${\mathcal{L}_{\delta}} = 20$ which is categorized as a VLES case. In this level and previous filter size, the DPSM and PSM models perform closely. Therefore, using dynamic model does not necessarily increase the performance. However, we observe the DNPS model still maintains the performance fairly high, according to the predictions of the peaks and tails of the distributions and correlation coefficients.   
\begin{figure}
    \begin{center}
        \begin{minipage}[b]{.49\linewidth}
            \centering
            \includegraphics[width=1\textwidth]{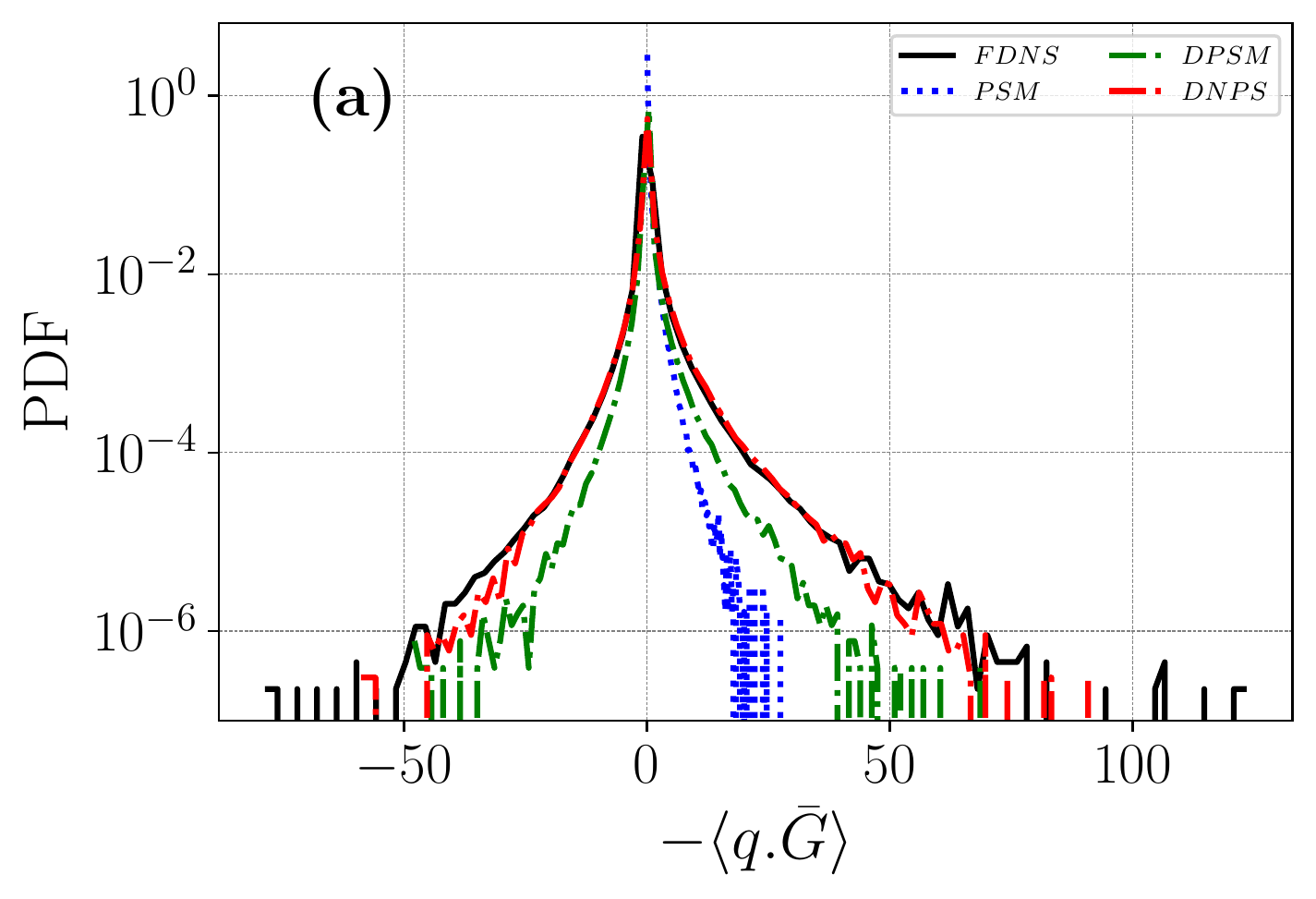}
        \end{minipage}
        \begin{minipage}[b]{.49\linewidth}
            \centering
            \includegraphics[width=1\textwidth]{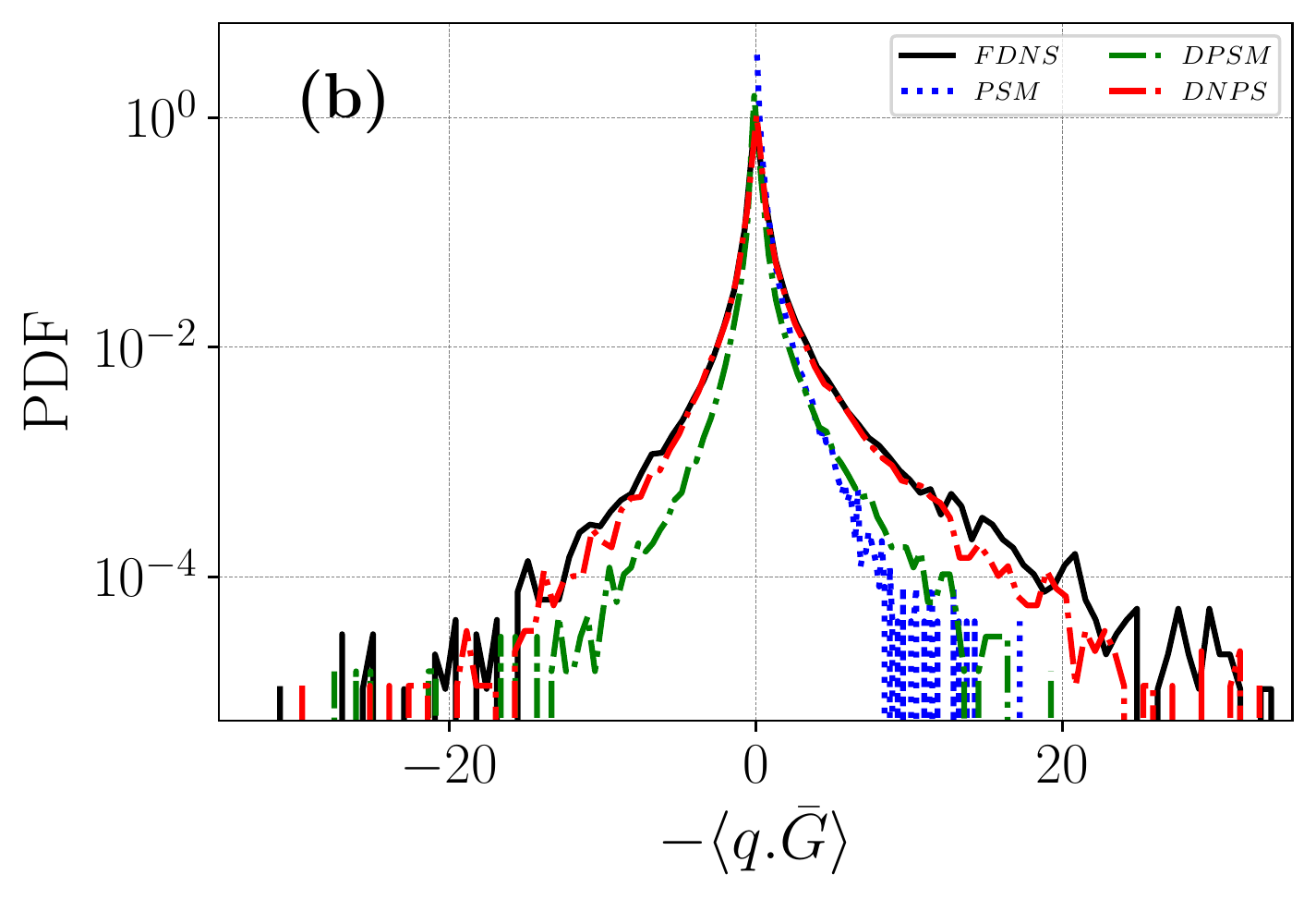}
        \end{minipage}
        \begin{minipage}[b]{.59\linewidth}
            \centering
            \includegraphics[width=1\textwidth]{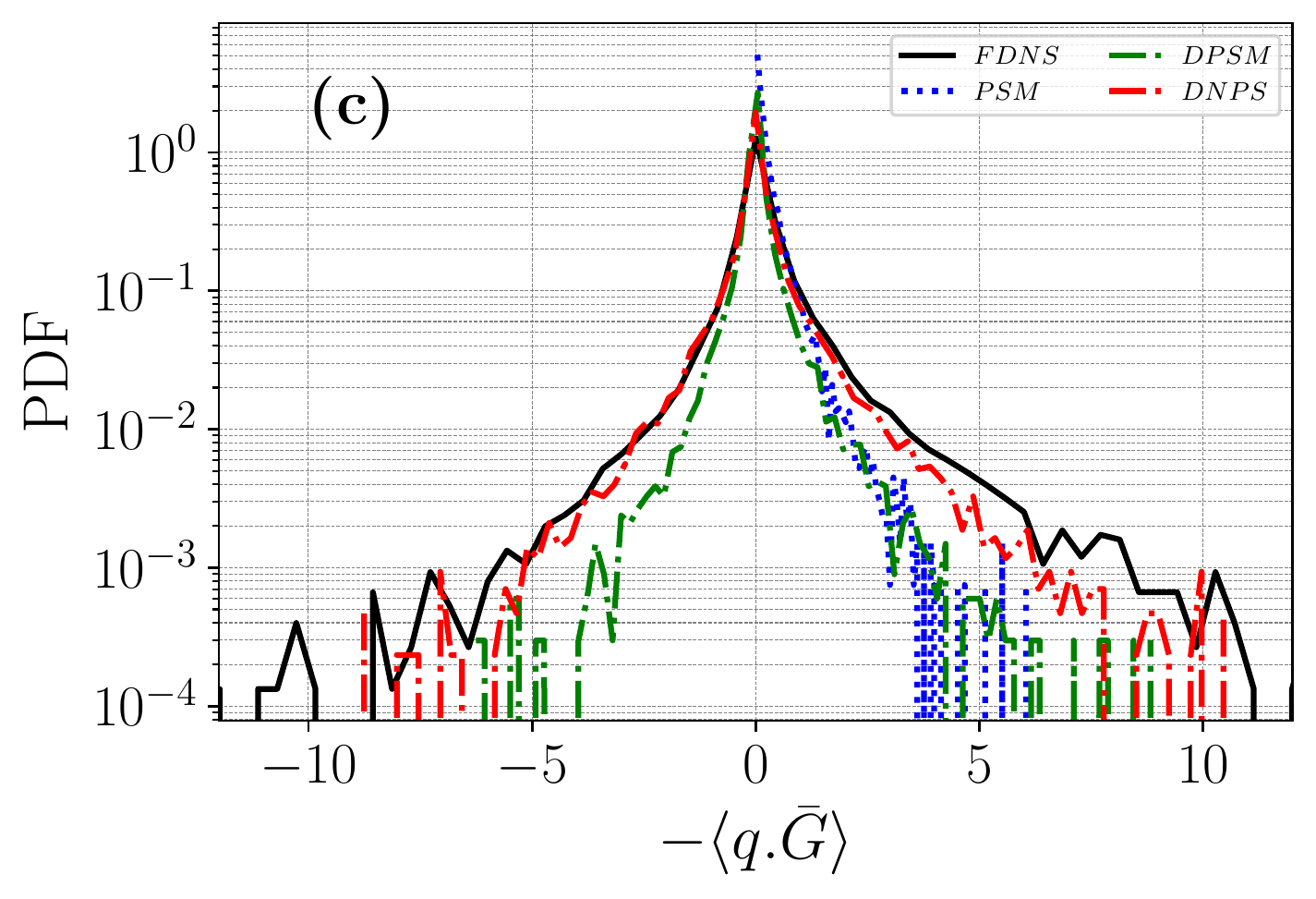}
        \end{minipage}
        \caption{\footnotesize  Normalized PDF of ensemble-averaged SGS dissipation using different models and different filter sizes, (a)  ${\mathcal{L}_{\delta}} = 4$ for LES, (b) ${\mathcal{L}_{\delta}} = 10$ for coarse LES, and (c) ${\mathcal{L}_{\delta}} = 20$ for VLES.}\label{fig: dissipation}
    \end{center}
\end{figure}

During the next test of the \textit{a priori} section, we assess the models' ability to predict the back-scattering phenomenon. Computing the SGS dissipation of the scalar variance provides a clear insight into the capability of the SGS models in reproducing the backward scattering in the turbulent cascade. This quantity is defined as 
\begin{equation}\label{Dissipation}
      \Pi = -\langle \boldsymbol{q}^R \cdot \bar{\boldsymbol{G}} \rangle,
\end{equation}
in which, $\bar{\boldsymbol{G}}$ represents the gradient of the filtered scalar field. Static models fail to capture the negative values for dissipation. However, the left wings do exist in the PDF of the ground-truth filtered DNS data, so it is part of the true physics. Furthermore, one of the major purposes of dynamic procedures is adding this important feature to the static models. Results in Figure \ref{fig: dissipation} illustrate the predicted dissipation using different models and filter sizes according to the above definition. As the three plots are showing, the PSM model fails to predict the left wings, back-scatter phenomena. DPSM model does predict the flow of scalar intensity from small scales to large scales, but the DNPS model has better agreement with the FDNS results. Moreover, in large filter sizes (${\mathcal{L}_{\delta}} = 10, 20$) we see that the DPSM model fails in predicting the peak of the PDF. Obtained results illustrate that dynamic nonlocal modeling not only adds the capability of back-scatter prediction to the static model, but also significantly increases its performance.%Considering the correlation coefficients between the values from ground-truth data and models at different filter sizes, we see that the DNPS model provides excellent estimations at all scenarios which are also visually obvious from the plots.
% =============================================================================
\section{\textit{A Posteriori} Tests}\label{sec: A Posteriori}
% =============================================================================
Evaluating the performance of any SGS model is ultimately targeted in an LES setting, where instead of utilizing the filtered DNS variables to construct the modeled closure terms, one can use the LES-\textit{resolved} flow variables, and apply the modeled closure for solving the LES equations through time. This method of assessment is called \textit{a posteriori} analysis which is coined by \citep{piomelli1988model} highlighting that the turbulence model is examined after being implemented in a numerical solver. Similar to the \textit{a priori} testing the reference values for comparisons are obtained from filtering the DNS-resolved flow variables. As a common practice in \textit{a posteriori} analysis of LES models, the time records of turbulent intensities are compared with their counterparts obtained from filtering the DNS results. For instance, in assessment of a model for the SGS scalar flux, resolved-scale scalar variance, $\frac{1}{2}\langle \fph^2 \rangle$, is the target turbulent intensity. Moreover, in more robust assessments, complex statistical quantities such as high-order structure functions of resolved-scale turbulent fields are compared with the ones obtained from filtered DNS \citep[see \textit{e.g.},][]{portwood2021interpreting, samiee2021tempered, seyedi2022data}. This type of examination, provides a sophisticated information on the two-point (indicating nonlocal correlations) and high-order statistical performance of the SGS model in an LES.

In order to perform the large-eddy simulations on the problem setting introduced in section \ref{sec: Gov-Eqns}, we employ the open-source pseudo-spectral solver developed in \citep{akhavan2020parallel}. We slightly modify this DNS framework to account for multiple SGS models of interest for $\boldsymbol{q}^R$, including the DNPS model we developed. This DNS framework has already been successfully utilized for the LES of decaying HIT flows with a variety of implemented models for the SGS stress tensor \citep{samiee2021tempered, seyedi2022data}. In an LES setting, both SGS stresses and fluxes in the filtered NS and AD equations are modeled, and a specific choice of SGS model for $\boldsymbol{\tau}^R$ could have a dominant or mixed effects on the resolved scalar concentration field \citep{portwood2021interpreting}. In our study, we choose to freeze these potential effects; therefore, we would be able to fully concentrate on the modeling aspects and performance for $\boldsymbol{q}^R$ in \eqref{eqn: F-AD-fluc}. As a result, in our numerical setup, we resolve the NS equations on DNS resolution, and throughout explicit filtering after each time-step we provide the velocity field on the LES resolution which is fed to the equation \eqref{eqn: F-AD-fluc}. This procedure has been employed in earlier studies such as \citep{portwood2021interpreting, vollant2016dynamic}.

According to the turbulent regime with $Re_\lambda=240$ and $Sc=1$ (as utilized in the calibration of $\alpha^{opt}$), we choose a fully-developed turbulent state for the velocity and scalar concentration from a well-resolved DNS. In order to reach to this turbulent state, the NS equations are resolved for approximately 15 $\tau_{LE}$ while an artificial forcing mechanism is enforced to the low wavenumbers (energy containing range) to maintain the turbulent kinetic energy \citep{akhavan2020parallel}. Afterwards, a we start resolving the AD equation from an initial fluctuating concentration of $\phi_0(\x)=0$ while imposing a uniform mean-gradient as described in section \ref{sec: Gov-Eqns}. By resolving the NS and AD equations for approximately 15 large-eddy turnover times, the skewness and flatness records of the fluctuating passive scalar gradient reaches to a statistically stationary state, ensuring the fully-developed turbulent scalar fluctuations are achieved. This procedure was successfully exercised in \citep{akhavan2020parallel, akhavan2021data, akhavan2021nonlocal}. By the explicit filtering of the velocity and scalar concentration fields, we take the initial condition for our LES tests. In our tests, we implement two static SGS models (PSM and NPS) in addition to two dynamic ones (DPSM and DNPS), and employ them in time-integration of filtered AD equation \eqref{eqn: F-AD-fluc} for 5 large-eddy turnover times. As mentioned in section \ref{subsec: SGS-modeling}, we utilized a stabilization process to avoid numerical instability. In this process, an averaging operator over the directions of statistical homogeneity is performed  \cite{germano1991dynamic}. Since our solver provides spectral accuracy in space and is not dissipative enough to handle the negative eddy-viscosity in long-time integration, we need to have this step in the implementation of the dynamic models in the \textit{a posteriori} assessments.
% =============================================================================
\subsection{Records of scalar variance}\label{subsec: Apost-variance}
~
\begin{figure}
    \begin{center}
        \begin{minipage}[b]{.49\linewidth}
            \centering
            \includegraphics[width=1\textwidth]{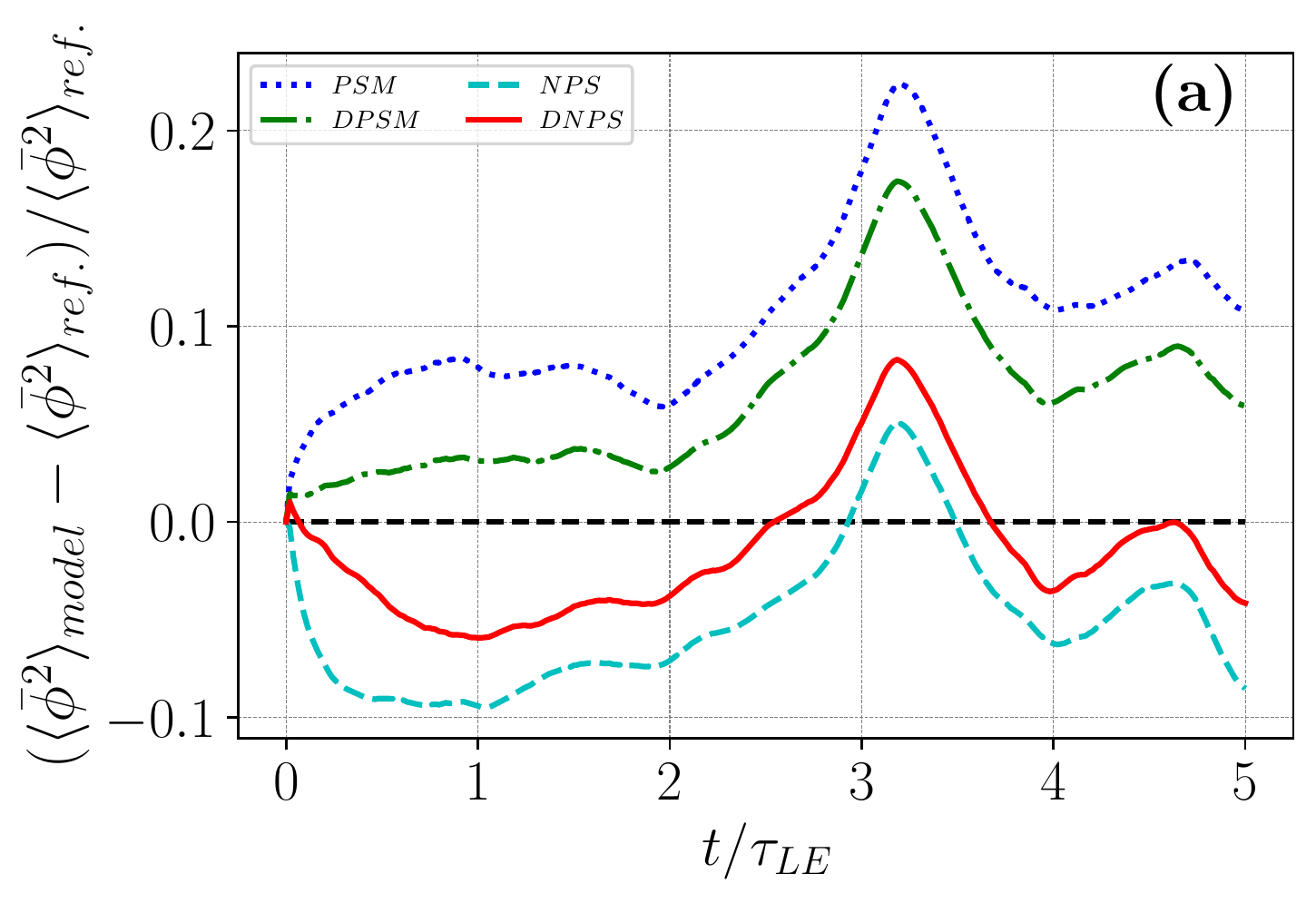}
        \end{minipage}
        \begin{minipage}[b]{.49\linewidth}
            \centering
            \includegraphics[width=1\textwidth]{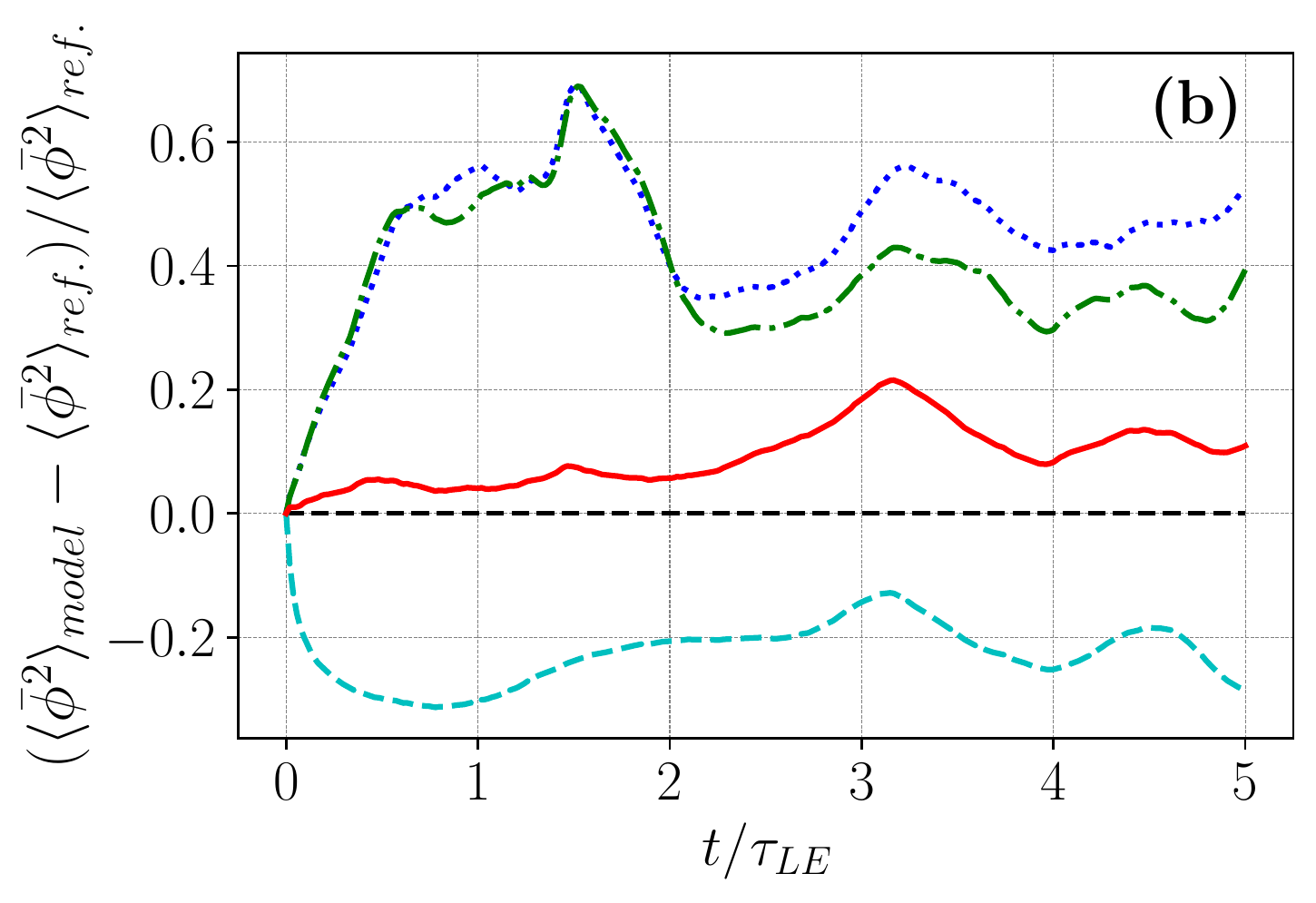}
        \end{minipage}
        \caption{\footnotesize  Relative error (with respect to the FDNS) in the records of scalar variance predicted in the LES for  (a): $\mathcal{L}_{\delta}=10$ and (b): $\mathcal{L}_{\delta}=20$.}\label{fig: Scalar_variance}
    \end{center}
\end{figure}

Evolution of the scalar variance is an important indicator in reliable prediction of the turbulent intensity. Figure \ref{fig: Scalar_variance}, shows the temporal records of relative error in the resolved-scale scalar variance using different SGS scalar flux models with respect to the obtained time record from filtering the DNS solution as the reference temporal record. The scalar variance errors are reported for the cases of coarse LES $(\mathcal{L}_{\delta}=10)$ and VLES $(\mathcal{L}_{\delta}=20)$ filter scale resolutions. The LES test cases as well as the reference DNS are conducted for 5 large-eddy turnover times.

For the large-eddy simulations on $\mathcal{L}_{\delta}=10$ resolution (Figure \ref{fig: Scalar_variance}a), we observe that at initial stage of the simulation $t/\tau_{LE}<2$, the DPSM model exhibits the lowest error compared to the other models in a way that on average over this time-span, the absolute value of the errors in the NPS, PSM, and DNPS models are 2.9, 2.5, and 1.4 times higher than the DPSM, respectively. However, for $2<t/\tau_{LE}$, the records of relative error indicate that nonlocal models (NPS and DNPS) start to perform better compared to the PSM and DPSM models. For instance, for the time-interval of $2.5 < t/\tau_{LE} < 5$, a comparison between the time-averaged errors from each model shows that the absolute value of error in the PSM, DPSM, and NPS models are 25, 18, and 7 times higher than what the DNPS model yields, respectively. In the LES tests with $\mathcal{L}_{\delta}=20$ (Figure \ref{fig: Scalar_variance}b), we observe approximately 9-10 times larger errors for the PSM and DPSM models compared to the DNPS model, when the records of error are averaged over the first 2 large-eddy turnover times of LES. This clearly shows that unlike the PSM and DPSM models, the DNPS model is reliably capable of keeping the modeling error at an acceptably low level while the LES is adjusting to the initial condition. For $2<t/\tau_{LE}<5$, the DPSM model exhibits lower errors compared to its non-dynamic version (approximately $25\%$ less error); however, DPSM model's error is approximately 3 times larger than the error recorded in the DNPS model. These numerical/statistical observations indicate that the dynamic procedure effectively improves the prediction of turbulent intensity in the LES at the long-time integration region. Moreover, they prove that the nonlocal models exhibit a remarkably better performance in the long-term prediction of resolved-scale scalar variance for different filter-scale resolutions.

% On the other hand, the nonlocal models (both the NPS and DNPS) provide better agreement with the ground-truth filtered DNS results (FDNS). Another important point that should be mentioned here is that the dynamic procedure is more gainful when it is leveraged in the nonlocal models rather than the conventional local models. The same story observed and discussed in \citep{seyedi2022data} for the prediction of the residual stresses. We focus on the VLES cases for the rest of the \textit{a posteriori} analysis where most of the conventional and local models fail to address correctly the true statistics. In the fine and Coarse LES regions ( $\mathcal{L}_{\delta}=4, 10$) we didn't see remarkable better performance for the DNPS model versus the DPSM. However, in the VLES region the performance of the new proposed model is noteworthy and surprising which opens door for the stable, efficient and fast turbulence modeling while preserving high-order structure functions. 

% =============================================================================
\subsection{Scalar structure functions and statistical nonlocality}\label{subsec: Apost-strs}
~
\begin{figure}
    \begin{center}
        \begin{minipage}[b]{.49\linewidth}
            \centering
            \includegraphics[width=1\textwidth]{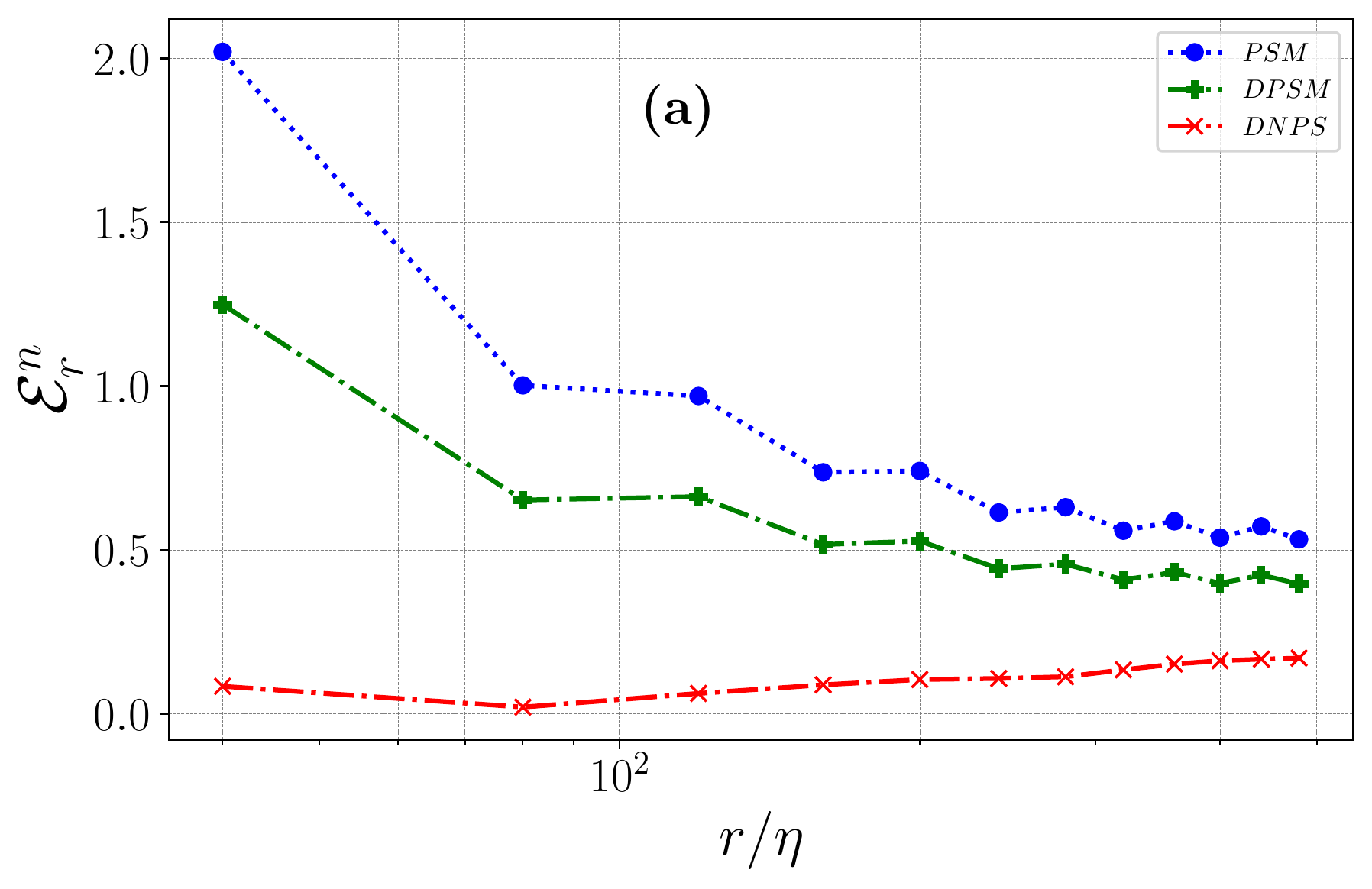}
        \end{minipage}
        \begin{minipage}[b]{.49\linewidth}
            \centering
            \includegraphics[width=1\textwidth]{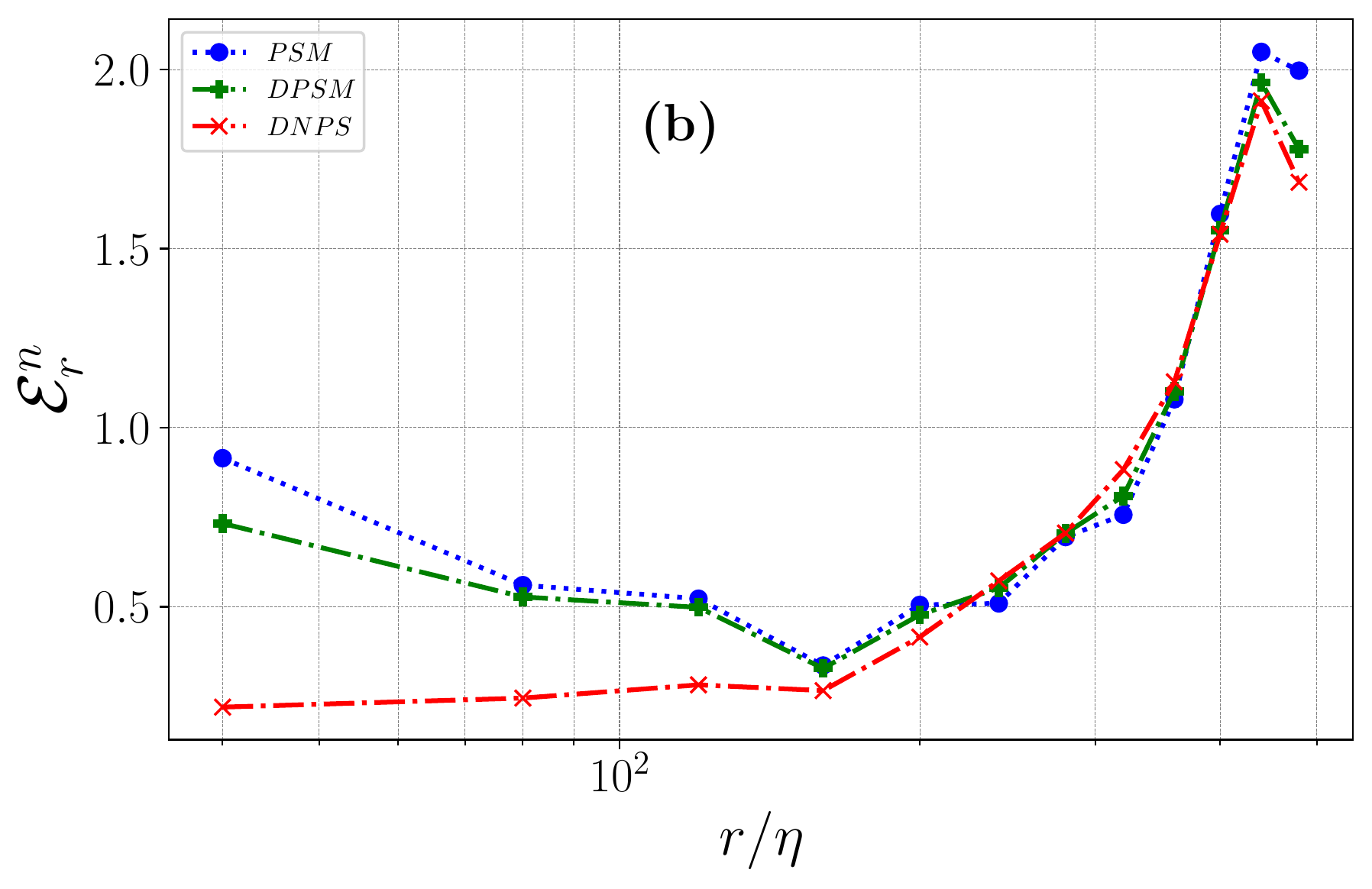}
        \end{minipage}
        \begin{minipage}[b]{.49\linewidth}
            \centering
            \includegraphics[width=1\textwidth]{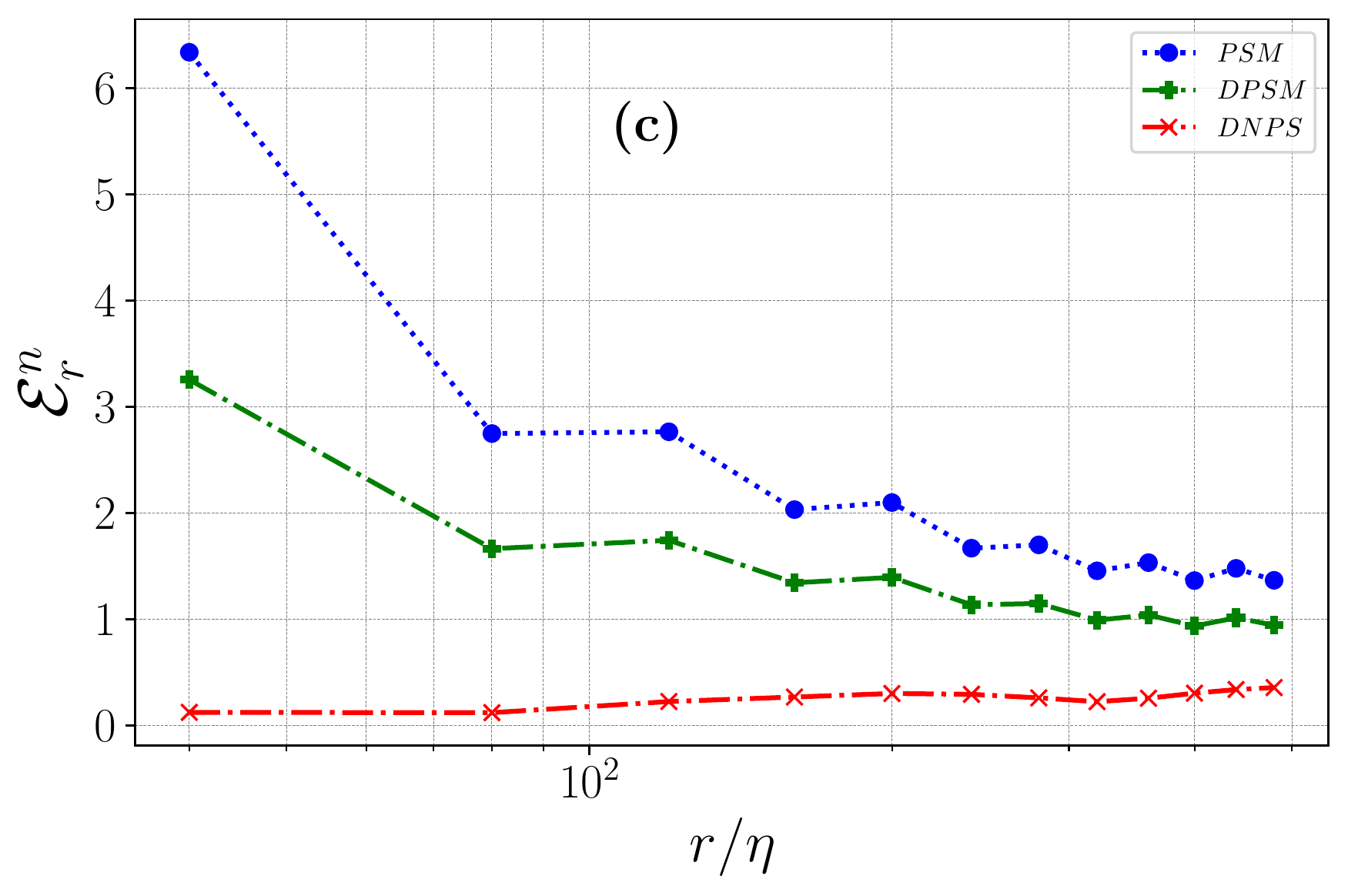}
        \end{minipage}
        \begin{minipage}[b]{.49\linewidth}
            \centering
            \includegraphics[width=1\textwidth]{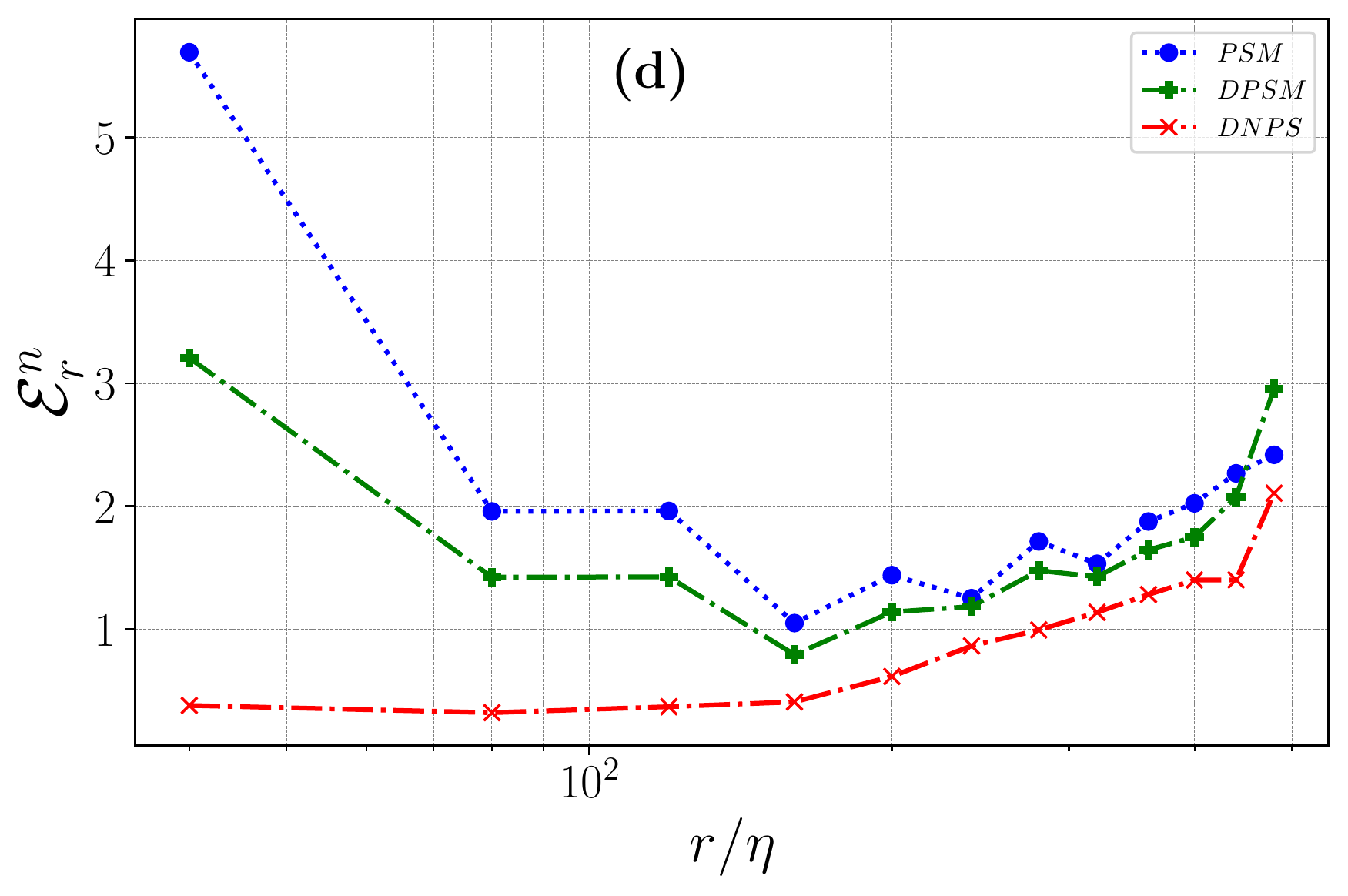}
        \end{minipage}

        \caption{\footnotesize  Time-averaged relative errors in the computed $\langle \delta_r \fph_L^n  \rangle$ from LES with different SGS models with respect to the FDNS as reference solution using $\mathcal{L}_{\delta}=20$. The time-averaging is done over $4<t/\tau_{LE}<5$ for (a): $n=2$, (b): $n=3$, (c): $n=4$, and (d): $n=5$.}\label{fig: Second_order}
    \end{center}
\end{figure}

Investigating the nonlocal behavior of turbulent regime is a vital and ultimate task in testing testing the performance of an SGS model in evolution of the turbulent field in LES \citep{meneveau1994statistics}. The structure functions of the resolved scalar field are robust two-point statistical measures that return $n$th-order statistics of resolved-scale scalar increments at a specific direction where $2 \leq n$ \citep{warhaft2000passive}. These structure functions of order $n$ are defined as
\begin{align}\label{eqn: str_func}
    \langle \delta_r \fph_L^n  \rangle =  \left\langle \big[ \fph_L (\x+r\boldsymbol{e}_L) - \fph_L (\x) \big]^{n}  \right\rangle; \quad n = 2,3,\dots, 
\end{align}
where $r$ is the size of spatial increment, $L$ represents the \textit{longitudinal} direction (the direction along the imposed uniform mean-gradient) \citep{iyer2014structure, portwood2021interpreting,  akhavan2021nonlocal}, and $\boldsymbol{e}_L$ specifies the unit vector along the longitudinal direction.

Considering the relative error between the $\langle \delta_r \fph_L^n  \rangle$ obtained from the LES solutions using an SGS model and the ground-truth FDNS solution for $\fph(\x)$ at a specific time, this error function is defined as:
\begin{align}\label{eqn: err_str_func}
    \mathcal{E}_r^n =  \left\vert \frac{\langle \delta_r \fph_L^n  \rangle_\mathrm{FDNS} - \langle \delta_r \fph_L^n  \rangle_\mathrm{LES}}{\langle \delta_r \fph_L^n  \rangle_\mathrm{FDNS} }\right\vert. 
\end{align}
Focusing on the temporal region $4 < t/\tau_{LE} \leq 5$ that the LES solution has undergone long time-integration, we select uniformly distributed samples of full-domain $\fph(\x)$ in time to compute $\mathcal{E}_r^n$ up to $n=5$. Since we are dealing with a statistically-stationary problem, we are allowed to take the temporal-average of these error function obtained from the sampled $n$th-order structure functions. Therefore, we have a robust indicator measure to examine the performance of each SGS model in predicting nonlocal and high-order statistics of resolved-scale scalar field in a long-time integrated LES. Figure \ref{fig: Second_order} illustrates this time-averaged $\mathcal{E}_r^n$ against the normalized spatial shift, $r/\eta$, for the PSM, DPSM, and the DNPS models. Figures \ref{fig: Second_order}a and \ref{fig: Second_order}c are showing the relative errors is the even-order structure functions, where the DNPS model has a considerably better performance. In particular, for the second-order structure functions (Figure \ref{fig: Second_order}a) the error for the DPSM model within $r/\eta<200$ region is considerably higher than what is observed for the DNPS model. In fact, for $r/\eta<100$ DPSM yields errors above 10 times greater than the reasonably small and steady errors we compute for the DNPS model, and for $100<r/\eta<200$ the DPSM errors are approximately 5 times larger. This important observation indicates significance of the dynamic nonlocal model in successful prediction of nonlocal statistics of $\fph$ (within the inertial-convective subrange) compared to its conventional counterpart (DPSM). Moreover, for $200<r/\eta$ we can still observe that the DNPS model's predictions for the time-averaged $\langle \delta_r \fph_L^2\rangle$ is still 2-3 time more accurate compared to the DPSM model. Interestingly, for the time-averaged $\mathcal{E}_r^4$ (Figure \ref{eqn: err_str_func}c) the similar behavior as described for the second-order structure function is observed supporting our argument on the effectiveness of the dynamic nonlocal model in prediction of high-order statistical nonlocality. On the other hand, the DNPS model exhibits better performance in predicting the odd-order $\langle \delta_r \fph_L^2\rangle$ compared to the DPSM model (see Figures \ref{eqn: err_str_func}b and \ref{eqn: err_str_func}d). In Figure \ref{eqn: err_str_func}b, it is observed that $\mathcal{E}_{r<200\eta}^3$ for the DNPS model is maintained at a reasonably low and steady level; however, the DPSM model returns up to 4 times larger errors. This remarkable performance of the DNPS model is even more highlighted when we look at the $\mathcal{E}_{r<200\eta}^5$ in Figure \ref{eqn: err_str_func}d, the DPSM model returns up to 9 times larger errors compared to its nonlocal counterpart. For $\mathcal{E}_{r>200\eta}^3$ (Figure \ref{eqn: err_str_func}b), the performance of both models are the same; however, by looking at $\mathcal{E}_{r>200\eta}^5$ (Figure \ref{eqn: err_str_func}d) the DNPS model always return lower errors. This comparison indicates that in prediction of the odd-order $\langle \delta_r \fph_L^n\rangle$, employing the DNPS model effectively improves the accuracy especially for the high-order ($3<n$) structure functions.

\section{Conclusions and Remarks}\label{sec: Conclusion}

% =============================================================================
In this paper, we developed a novel dynamic nonlocal closure model for the subgrid-scale scalar field in the context of the large and very large eddy simulation (LES, VLES). With our high-fidelity datasets pertaining to forced homogeneous isotropic turbulence, we examined the effect of fractional order and characteristic filter sizes in LES and VLES cases. During enough large-eddy turnover times, we utilized ensemble-averaged quantities from ten separate three-dimensional snapshots to make final decisions. When the ground-truth force and the predicted SGS force were most correlated, the optimal fractional order was selected for each scenario. We initially tested the proposed model in the context of \textit{a priori} assessment. Based on the results, we showed that the new DNPS model is more accurate in predicting SGS dissipation and force terms. Furthermore, the DNPS model retained a fair performance in the VLES cases, unlike other conventional models that did not return accurate prediction of LES closure at very large filter sizes.
In an LES setting, we managed to examine the performance of SGS models in an \textit{a posteriori} sense. We resolved the filtered AD equation for 5 large-eddy turnover times while $\fU$ contribution in the advective coupling obtained from the explicit filtering of the DNS solution. We looked at the relative error (with respect to the FDNS) in the records of the $\langle \fph^2 \rangle$ from the LES solution using different SGS models. Tracking these records showed that the combination of nonlocal modeling and dynamic procedure (DNPS modeling) is an effective way for accurate prediction of the resolved-scale turbulent intensity (scalar variance) especially when the goal is to study the long-term behavior. Moreover, we examined the prediction of the longitudinal resolved-scale scalar structure functions, $\langle \delta_r\fph^n_L\rangle$, for $2 \leq n \leq 5$. Compared to their time-averaged FDNS solution over $4 \leq t/\tau_{LE} \leq 5$, we observed that the time-averaged LES solution obtained from utilizing the DNPS model is performing remarkably successful in maintaining a low-level error over the multitude of scales for spatial shift. Therefore, we realized that unlike the PSM and DPSM model, the DNPS model does a great job in prediction of the high-order and two-point (nonlocal) statistics of $\fph$ especially in over the inertial-convective subrange. 

In conclusion, we showed that employing larger filter sizes instead of conventional LES filter sizes in the proposed model is a promising direction. In this way, the high computational costs associated with fractional modeling can be compensated, and one can achieve stable, reasonably accurate, and fast simulations using the new closure model.

% =============================================================================
%                Acknowledgement
% =============================================================================
\section*{Acknowledgement}
This material is based upon work supported by the MURI/ARO award (W911NF-15-1-0562), the AFOSR Young Investigator Program (YIP) award (FA9550-17-1-0150), the ARO YIP award (W911NF-19-1-0444), and the NSF award (DMS-1923201). The HPC resources and services were provided by the Institute for Cyber-Enabled Research (ICER) at Michigan State University.
% =============================================================================
%                Appendix
% =============================================================================
\appendix

\section{Mathematical concepts utilized in the model derivation}\label{sec: Appendix1}
In this section, we provide mathematical definitions and concepts that have been used in the model development sections. The Fourier transform of the fractional Laplacian can be obtained using the below relation on the unbounded domains \cite{lischke2020fractional} as
\begin{eqnarray}\label{eq:Fourier}
          \mathcal{F} \Big [ (-\Delta) ^ {\alpha} u(x) \Big] = |\k|^{2\alpha} \mathcal{F} [u] (\k).
\end{eqnarray}
In this equation $(-\Delta)^{\alpha}$ (with $0 < \alpha \leq 1$) denotes the fractional operator and $\mathcal{F}$, $\k$ show the Fourier transform and Fourier wavenumber vector, respectively. Setting $\alpha = 1$ will result to integer-order Laplacian. Moreover, the definition of $\alpha$-Riesz potential is
\begin{eqnarray}\label{FL-3-2}
    \mathcal{I}_{\alpha} u(\boldsymbol{x}) &=& C_{d,-\alpha} \, \int_{\mathbb{R}^d }\frac{u(\boldsymbol{x})-u(\boldsymbol{s})}{\vert \boldsymbol{x}-\boldsymbol{s}\vert^{d-2\alpha}} \, d\boldsymbol{s},
\end{eqnarray}
the fractional Laplacian operator can also be expressed in the integral form using the mentioned definition as
\begin{eqnarray}\label{FL-3}
    (-\Delta)^{\alpha} u(\boldsymbol{x}) &=& C_{d,\alpha} \, \int_{\mathbb{R}^d }\frac{u(\boldsymbol{x})-u(\boldsymbol{s})}{\vert \boldsymbol{x}-\boldsymbol{s}\vert^{2\alpha+d}} \, d\boldsymbol{s},
\end{eqnarray}
in which $C_{d,\alpha} = \frac{2^{2\alpha} \Gamma(\alpha+d/2)}{\pi^{d/2} \Gamma(-\alpha)}$ for $\alpha \in (0,1]$ and $\Gamma(\cdot)$ is the Gamma function \citep{lischke2020fractional}. The $\alpha$-Riesz potential can also be written as \citep{stein2016singular}
\begin{eqnarray}\label{FL-3-3}
    \mathcal{I}_{\alpha} u(\boldsymbol{x}) &=& (-\Delta)^{-\alpha} u(\boldsymbol{x}) = \mathcal{F}^{-1} \Big {\{}   \vert \boldsymbol{\k} \vert^{-2\alpha} \mathcal{F}  \big {\{}u\big {\}} (\boldsymbol{\k}) \Big {\}}.
\end{eqnarray}
The Riesz transform is derived given \eqref{FL-3-3} 
\begin{eqnarray}\label{FL-3-4}
    \boldsymbol{\mathcal{R}}_j u(\boldsymbol{x}) &=&\nabla_j \, \mathcal{I}_{1} u(\boldsymbol{x}) = \mathcal{F}^{-1} \Big {\{}  -\mathfrak{i} \, \frac{k_j}{ \vert \boldsymbol{\k} \vert} \mathcal{F}  \big {\{}u\big {\}} (\boldsymbol{\k}) \Big {\}},
\end{eqnarray}
which is utilized in formulating the dynamic nonlocal SGS model.

% =============================================================================
% =============================================================================================
\bibliographystyle{jfm}
\bibliography{mybib}

\end{document}